\documentclass[twocolumn,aps,prb,amsmath,showpacs]{revtex4}
\usepackage{graphicx,dcolumn,bm,amssymb,amsmath}
\usepackage{color}

\def\be{\begin{equation}}
\def\ee{\end{equation}}

\begin{document}
\author{M. Schlegel$^1$, J. Brujic$^2$, E. M. Terentjev$^{3}$ and A.
Zaccone$^{4}$}
\affiliation{${}^1$Department of Engineering, University of Cambridge, Trumpington Street, CB2 1PZ Cambridge,
U.K.}
\affiliation{${}^2$Physics Department, New York University, New York, NY 10003,
USA}
\affiliation{${}^3$Cavendish Laboratory, University of Cambridge, JJ Thomson Avenue, CB3 0HE Cambridge,
U.K.}
\affiliation{${}^4$Department of Chemical Engineering and Biotechnology, University of Cambridge, New Museums Site, Pembroke Street, CB2 3RA Cambridge, U.K.}

\begin{abstract}
\noindent
Paradigmatic model systems, which are used to study the mechanical response of
matter,
are random networks of point-atoms, random sphere packings, or simple crystal
lattices; all of these models assume central-force interactions between
particles/atoms. Each of these models differs in the spatial arrangement and
the correlations among particles. In turn, this is reflected in the widely
different behaviours of the shear ($G$) and compression ($K$) elastic moduli.
The relation between the macroscopic elasticity as encoded in $G$, $K$ and
their ratio, and the microscopic lattice structure/order, is not understood.
We provide a quantitative analytical connection between the local orientational
order and the elasticity in model amorphous solids with different internal
microstructure, focusing on the two opposite limits of packings (strong
excluded-volume) and networks (no excluded-volume).
The theory predicts that, in packings, the local orientational order due to
excluded-volume causes less nonaffinity (less softness or larger stiffness)
under compression than under shear. This leads to lower values of $G/K$, a
well-documented phenomenon which was lacking a microscopic explanation. The theory
also provides an excellent one-parameter description of the elasticity of
compressed emulsions in comparison with experimental data over a broad range of
packing fractions.

\end{abstract}

\title{Local structure controls shear and bulk moduli in disordered solids}
\maketitle

One of the overarching goals of solid state physics is to find a universal
relationships between the lattice structure of matter in the solid state and
its mechanical response. From this point of view, it is important to simplify
the details of the
interactions between the building blocks (atoms, particles) in order to single
out the relevant physics
and general laws.
The framework of lattice dynamics successfully provided the link between
atomic-level
structure and macroscopic properties of simple crystal
lattices~\cite{Born1954}. Our understanding is instead much more limited
when structural disorder plays an important role, such as in glasses, liquids
and other disordered states of matter~\cite{goodrich,amir}.

With the advent of computer simulations, it became clear that disordered
solids, which are of paramount importance in many areas of technology and life
sciences,
cannot be described simply as perturbations about the crystalline order.
In this context, an unsolved problem is the striking difference in the elastic
deformation behaviour of random networks and random packings. For
networks, the shear modulus $G$ and the compression modulus $K$ display the
same dependence on the coordination number $z$ which represents the average
number of elastic springs per node of the network. Therefore, $G\propto K
\propto (z-z_c)$, and both moduli vanish at the same critical coordination
$z_c$ which is dictated by isostaticity. It is different for random packings
where only the shear modulus scales \textit{linearly} as $G\propto (z-z_c)$, whereas the bulk
modulus vanishes only at a coordination much lower than $z_c$. This means that
packings have a comparatively larger bulk modulus, with respect to random
networks, and remain well stable against compression also near, at, and  even
below the critical coordination where shear rigidity vanishes. This state of
affairs has been revealed in simulation
studies~\cite{ellenbroek2009non,goodrich}, at least since the
1970's~\cite{weaire}. Furthermore, the same phenomenon is well documented also
in disordered atomic solids~\cite{Kinder} and non-centrosymmetric crystals
(e.g. piezoelectrics) ~\cite{Helmreich}.

However, there is no mechanistic understanding of this phenomenon, nor
analytical theories able to describe it, beyond the somewhat obvious
observation that the internal structure of packings is different from that of
random networks, due to the self-organization and mutual excluded-volume of
particles in the packing, which are absent in random isotropic networks. Below
we provide a quantitative connection between structure and elasticity based on
nonaffine lattice dynamics which shows that the local self-organization of the
particles with excluded-volume leads to a higher degree of bond-orientational
order~\cite{Tanaka2010,Tanaka2013} in randomly packed structures compared to
isotropic random networks. In turn, this leads to a significantly higher bulk
modulus and a lower nonaffinity under compression.\\

\textbf{Results}

\textbf{Nonaffine lattice dynamics.}
Our main tool is the Born-Huang free energy expansion~\cite{Alexander},
suitably
modified to account for the structural disorder in terms of the nonaffinity of
the displacements (as explained below). In order to make analytical
calculations, we neglect the
effect of thermal fluctuations (i.e. we operate in the athermal limit, which is
applicable to granular solids and non-Brownian emulsions), and we focus on
harmonic central-force interactions between the particles. Thus we neglect
both the bending resistance when the particles slide past each other, as well
as the effect of stressed bonds.
It is important to emphasize that both these effects can provide rigidity to
certain lattices, which are otherwise floppy or unstable when only central
forces between atoms are active. This fact is well known e.g. in the context of
inorganic network glasses~\cite{Thorpe, Alexander}.

The key to understanding the elasticity of amorphous lattices is
nonaffinity~\cite{Alexander}. In a nutshell: the applied
external deformation induces a deformation at the microscopic level of
interatomic bonds. If the interatomic displacements are simply proportional to
the applied overall deformation field, then the deformation is called
\emph{affine}, and one can expand the free energy in powers of small
interatomic
displacements and take the continuum limit of the microscopic deformation for
either shear deformation or
compression~\cite{Born1954}. In other words, the microscopic interparticle
displacements
are directly proportional to, and uniquely determined by the applied
macroscopic strain.
Differentiating the free energy twice with respect to the macroscopic strain
yields the
shear modulus $G$ and the bulk modulus $K$, depending on the geometry of the
applied deformation (shear or hydrostatic compression, respectively).

As was first realized by Lord Kelvin~\cite{Kelvin}, and more recently
emphasized by Alexander~\cite{Alexander} and Lemaitre and
Maloney~\cite{Lemaitre}, the affine approximation is strictly valid only for
centrosymmetric crystal lattices. The reason becomes evident if one considers
the forces which are transmitted to a test atom in the lattice upon deforming
the
solid. Every neighbour transmits a force which is cancelled by the
\textit{local} inversion symmetry
in the centrosymmetric Bravais cell (see Fig.1a below). As a result, there is
no
local net force acting on the atoms of the lattice in their affine positions,
and the old affine free energy
expansion~\cite{Born1954} suffices to correctly describe the elastic
deformation. With a disordered or non-centrosymmetric lattice, the situation is
different. The
forces that every atom receives from its neighbours no longer cancel, because
the local inversion symmetry is violated. The net force acting on every atom
has to
be relaxed via additional atomic displacements, called \emph{nonaffine}
displacements~\cite{Lemaitre}. These motions, under the action of the
disorder-induced local forces, are associated with a total work, which is an
internal work done by the system (hence negative, by thermodynamic convention).

The work done by nonaffine displacements represents a quote of internal lattice
energy which cannot be employed
to react to the applied deformation. Therefore, the free energy
of deformation can be written as $F=F_\mathrm{A}-F_\mathrm{NA}$, to distinguish
the affine contribution $F_\mathrm{A}$ from the nonaffine contribution
due to disorder~\cite{zaccone2011approximate,zaccone2011approximate2},
$-F_\mathrm{NA}$.
The fact that non-centrosymmetric lattices (e.g. piezoelectric crystals) are
affected by nonaffine distortions of the primitive cell~\cite{Elliott, Tilley},
however, does not necessarily mean that
they are unstable or soft. These materials are, of course, fully rigid and do
exhibit a large value of shear modulus, provided that they have a sufficient
atomic coordination, well above the isostatic limit, and a fairly large value
of spring constants.

\textbf{Theory of elastic moduli.}
Upon carrying out the formal treatment with the standard dynamical (Hessian)
matrix~\cite{Ashcroft} $\underline {\underline{H}}_{ij}$ and the expression for
the disorder-induced force (defined for the example of shear deformation
$\gamma$ in the $\{xy\}$ plane as
$\underline{f}_{i}=\underline{\Xi }_{i}^{xy}\gamma$), the nonaffine
contribution
to the free energy of deformation can be evaluated as shown in several places
in the recent literature~\cite{Lemaitre,zaccone2011approximate}. It has been
shown that the elastic constants are given by
$C_{\iota \xi \kappa \chi }=C_{\iota \xi \kappa \chi }^\mathrm{A}-C_{\iota \xi
\kappa \chi }^\mathrm{NA}$ with the nonaffine correction due to disorder given
as
\begin{equation}
C_{\iota \xi \kappa \chi }^\mathrm{NA}=\sum_{ij}\underline{\Xi }_{i}^{\iota
\xi }\left( \underline {\underline{H}}_{ij}\right)^{-1}\underline{\Xi
}_{j}^{\kappa \chi  }.
\end{equation}
The affine part of the elastic constants is provided by the affine Born-Huang
lattice dynamics, which is exact for centrosymmetric lattices:  $C_{\iota \xi
\kappa \chi }^\mathrm{A}= \frac{1}{2V} R_{0}^{2}\kappa
\sum_{ij}n_{ij}^{\iota}n_{ij}^{\kappa}n_{ij}^{\xi}n_{ij}^{\chi}$.
Here $\kappa$ is the effective spring constant of the interatomic
(interparticle) interaction, which is harmonic near the equilibrium, $V$ is the
total volume of the system, and $R_{0}$ is the equilibrium separation length
between nearest neighbours spheres of diameter $\sigma$. $n_{ij}^{\iota}$ is
the $\iota=x,y,z$ Cartesian
coordinate of the unit vector which defines the orientation of the bond between
two bonded neighbours $i$ and $j$.
In the nonaffine relaxation term, the force per unit strain acting on every
atom is given analytically, for the case of shear deformation,
by~\cite{Lemaitre}
$\underline{\Xi }_{i}^{xy}=-R_{0}\kappa
\sum_{j}\underline{n}_{ij}n_{ij}^{x}n_{ij}^{y}$. It is easy  to check that
$\underline{\Xi }_{i}^{xy}=0$ for a centrosymmetric lattices.
As shown in Ref.\cite{zaccone2011approximate}, under the assumption of
central-force interaction, and for a random network of
equal harmonic springs with number density of nodes $N/V$, the shear modulus
can be evaluated analytically as
\begin{align}\label{nonaffine}
        G=G_A-G_{NA}=\dfrac{1}{30}\frac{N}{V} \kappa R_0^{2}  (z-z_c).
\end{align}
The proportionality to $z$ is contributed by the affine term $C_{xyxy}^\mathrm{A}$ above, where the
sum $\sum_{ij}n_{ij}^{\iota}n_{ij}^{\kappa}n_{ij}^{\xi}n_{ij}^{\chi}$ can be evaluated in mean-field 
averaging, $(1/2)\sum_{ij}n_{ij}^{x}n_{ij}^{y}n_{ij}^{x}n_{ij}^{y}\simeq (zN/2)\langle n_{ij}^{x}n_{ij}^{y}n_{ij}^{x}n_{ij}^{y} \rangle$, where 
the quantity $(zN/2)$ represents the total number of bonds in the system. The factor $1/2$ in front of the $\sum_{ij}...$ is required because the sum counts the
bonds twice. Further, $\langle n_{ij}^{x}n_{ij}^{y}n_{ij}^{x}n_{ij}^{y} \rangle=1/15$ for a random isotropic distribution of bond orientations.

The nonaffinity of the amorphous solid is encoded in the quantity $C_{\iota \xi
\kappa \chi }^\mathrm{NA}\propto -z_c$, which defines the critical number $z_{c}=2d=6$ of
bonds at which the shear modulus vanishes by  virtue of the non-affine
softening mechanism.

\begin{figure*}[t]
\centering
{\includegraphics[width=0.95\textwidth]{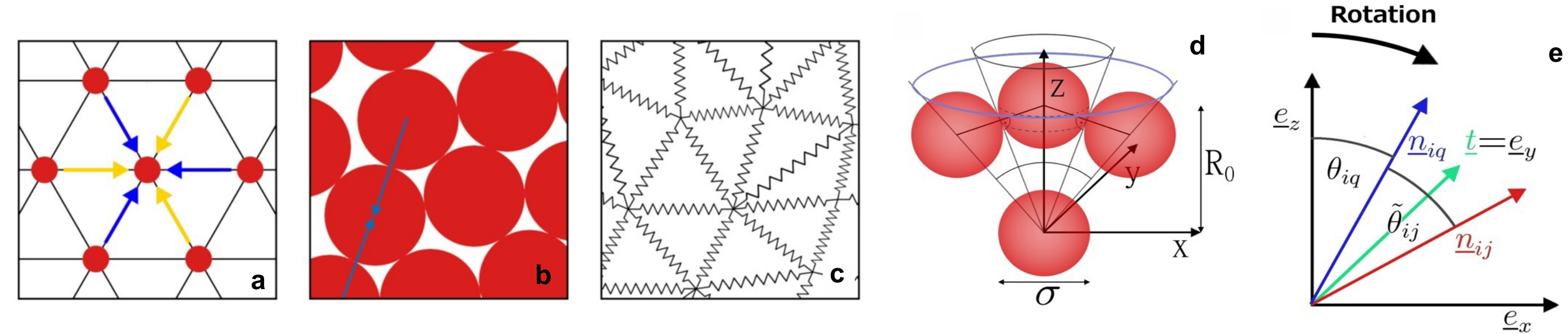}
\caption{\textbf{Geometry of particles, bonds and forces:} \ (a) In a
centrosymmetric lattice the forces acting on every particle cancel by symmetry
and leave the particle force-free. Hence no additional displacements are
required to keep local mechanical equilibrium on top of the affine
displacements dictated by the applied strain. (b) In a jammed packing, there is
a remarkable degree of local orientational order: due to excluded-volume
correlations it can
still happen that two particles make an angle equal to $180^o$ across the common
neighbour at
the center of the frame, leading to cancellation of local forces. This
effect is significant under compression, thanks to isotropy, but
negligible under shear. (c) In a random network, the probability of having
this cancellation of forces is much smaller. In this case, nonaffine
displacements are
required on all particles (nodes) to keep local equilibrium under the non-vanishing sum of nearest-neighbour forces. 
This limit has the
strongest nonaffinity and the lowest values of elastic moduli. (d) The excluded
volume cone: a bond, for example along the $z$-axis, leads to an excluded-cone
where no third particle can exist. $R_{0}$ is the equilibrium bond distance,
$\sigma$ represents the diameter of the particles. (e) The frame-rotation trick
to evaluate the contributions of local excluded-volume correlations to the
nonaffine elastic moduli. Here, for simplicity, only the special case of $\phi_{ij}=\phi_{iq}=0$, i.e. both $ij$ and $iq$ lying in the plane $xz$, has been illustrated. }}
\end{figure*}

This result is valid for random networks where bonds have randomly distributed
orientations in the solid angle.
In that model, any bond-orientational order parameter is identically zero and
the average rotational symmetry is isotropic.
For the more general case where correlations between bond-orientation vectors
of nearest-neighbours are important, it  can be shown (see the Supplementary
Information) that the nonaffine correction term reduces to the following
form, after replacing the sum over bonds by the average:
\begin{equation}
C_{\iota\xi\kappa\chi}^{NA}= \kappa
R_{0}^{2}3\frac{N}{V}\sum_{\alpha=x,y,z}\left(A_{\alpha,\iota\xi\kappa\chi}+
B_{\alpha,\iota\xi\kappa\chi}\right ) ,
\end{equation}
where $A_{\alpha,\iota \xi \kappa \chi }\leq0$ and $B_{\alpha,\iota \xi \kappa
\chi } \geq0$ are defined as follows:
\begin{subequations}
\begin{align}
A_{\alpha,\iota \xi \kappa \chi }&=\left \langle
n_{ij}^{\alpha}n_{ij}^{\iota}n_{ij}^{\xi
}n_{iq}^{\alpha}n_{iq}^{\kappa}n_{iq}^{\chi} \right \rangle\\
B_{\alpha,\iota \xi \kappa \chi }&=\left \langle
n_{ij}^{\alpha}n_{ij}^{\iota}n_{ij}^{\xi
}n_{ij}^{\alpha}n_{ij}^{\kappa}n_{ij}^{\chi} \right \rangle.
\end{align}
\end{subequations}
Here $\langle... \rangle$ represents an angular average, in the solid angle,
over the orientations of bonds $ij$ and $iq$ as explained
in~\cite{zaccone2011approximate2}. It is important to note that in
$A_{\alpha,\iota \xi \kappa \chi }$, we average over all possible orientations
of two bonds to the atoms $j$ and $q$, respectively, measured from a common
atom $i$. For the average in $B_{\alpha,\iota \xi \kappa \chi}$, one only needs
to consider bonds between the particles $i$ and $j$ as discussed
in~\cite{zaccone2011approximate}. Hence, it is evident that  $A_{\alpha,\iota
\xi \kappa \chi}$ is non-zero only if the orientations of the two bonds $ij$
and $iq$ are correlated (that is, the orientation of $ij$ does depend on the
orientation of $iq$, and vice versa). If there is no correlation, meaning that
given a certain orientation of $iq$ in the solid angle, $ij$ can have any
random orientation in the solid angle with the same probability, then
$A_{\alpha,\iota \xi \kappa \chi} =0$. This is so because the average can be
factored out into the product of two averages of triplets each of the type
$\langle n_{ij}^{\alpha}n_{ij}^{\iota}n_{ij}^{\xi }\rangle$, and each angular
average vanishes separately, as one can verify by insertion.

The limit where any two bonds $ij$ and
$iq$ are uncorrelated, and $A_{\alpha,\iota\xi\kappa\chi}=0$, defines the
geometry of the random
network~\cite{ellenbroek2009non} (Fig.1c). The random network limit represents
the case where nonaffinity makes the largest negative correction to the elastic
constants, thus softening the material. The random network is thus the opposite
extreme to the perfect centrosymmetric hard crystal.

In the random network model, which served for long time as a structural model
for
many inorganic glasses~\cite{Thorpe,Boolchand}, the nodes are just point-atoms
with zero volume, $\sigma=0$. This is a very important feature because the
absence of any
excluded-volume hindrance between such atoms allows them to be placed
at random positions in space. Such a model is clearly applicable only to
systems where
the bond length is much larger than the atomic diameter $\sigma$ (which is the
case for network glasses and some amorphous semiconductors). The limit
$\sigma/R_{0} \rightarrow 0$ thus corresponds to the random network model.  The
opposite limit, $\sigma/R_{0} = 1$, corresponds to the jammed
packing, where spherical particles are barely touching their neighbours.
In this limit, the excluded-volume repulsion between spheres in close contact
plays a very important role in the self-organization and in the local
structure of the packing. In particular,
due to excluded-volume, there are restrictions on the available portion of
solid angle where a nearest-neighbour can sit. It is therefore significantly
more likely, in comparison with the random network case, that a particle $j$
makes an angle of $180^{o}$ with a particle $q$ directly across a third
particle $i$ placed at the center of the frame (Fig.1b),
due to the existence of sectors in the solid angle (as measured from the
central particle) that are forbidden. Hence, the local orientational order in
the jammed
packing, well documented in previous structural
studies~\cite{Tanaka2010,Tanaka2013}, is important
also in the determination of elastic moduli.
In the following we are going to focus our detailed calculations on the jammed
packing limit with $R_{0}=\sigma$.

We implemented a minimal model, inspired by the granocentric model of
granular packings~\cite{granocentric}, for the excluded-volume correlations
which allows an explicit evaluation of the two-bond angular-correlation terms
$A_{\alpha,\iota \xi \kappa \chi }$ for jammed packings. If the bond $iq$ has a
given orientation in the solid angle, parameterised by the pair of angles
$\{\varphi_{iq},\theta_{iq}\}$ then, clearly, the bond $ij$ can have any
orientation in the solid angle apart from those orientations delimited by the
excluded cone depicted in Fig.1d. The angular average for the orientation of
$ij$ is thus restricted to the total solid angle $\Omega$ minus the
excluded cone, which gives 
the allowed solid angle as $\Omega-\Omega_\mathrm{cone} $, with
$\Omega_\mathrm{cone}=\pi(\sigma/R_{0})^{2}$. The probability density
distribution $\rho$ of bond orientations is taken to be isotropic for
$iq$,
that is $\rho_{iq}=1/4\pi$. For $ij$, instead, the probability that it takes a certain orientation is a conditional one, because it depends on the orientation of $iq$.
Hence, the conditional probability for the orientation of $ij$ 
is $\rho_{ij}(\Omega_{ij}\mid\Omega_{iq})=1/(4\pi-\Omega_\mathrm{cone})$, for $\Omega_{ij}\in {\Omega-\Omega_\mathrm{cone}}$, and $\rho_{ij}(\Omega_{ij}\mid\Omega_{iq})=0$ for $\Omega_{ij}\in {\Omega_\mathrm{cone}}$. In the section below we use these considerations
to evaluate the excluded-volume correction to the nonaffine moduli encoded in $A_{\alpha,\iota\xi\kappa\chi}$.

\textbf{Evaluation of the excluded-volume correlations term in the moduli}.
The excluded-volume correlation term contributing to the elastic moduli is given by
\begin{subequations}
\begin{align}
A_{\alpha,\iota \xi \kappa \chi }&=\left \langle
n_{ij}^{\alpha}n_{ij}^{\iota}n_{ij}^{\xi
}n_{iq}^{\alpha}n_{iq}^{\kappa}n_{iq}^{\chi} \right \rangle\\
&\begin{aligned}
[t]
=&\int_{\Omega}\int_{\Omega-\Omega_{\mathrm{cone}}}\rho_{ij}(\Omega_{ij}\mid\Omega_{iq})\rho_{iq}\left(\Omega_{iq}\right)
n_{iq}^{\alpha}n_{iq}^{\kappa}n_{iq}^{\chi} \\
& \times
n_{ij}^{\alpha}n_{ij}^{\iota}n_{ij}^{\xi
}\textup{d}\Omega_{ij}\textup{d}\Omega_{iq}.
\end{aligned}
\end{align}
\end{subequations}
To evaluate the above integral it is necessary to first identify the
correlation between $ij$ and $iq$ and then devise a strategy to evaluate the
integral in the above equation.

A solution can be found by exploiting the symmetry of the problem,
and, in particular, the rotational invariance. The local Cartesian frame
centered on the particle $i$ is rotated such that the $z$-axis (from which the
azimuthal angles $\theta_{ij}$ and $\theta_{iq}$ are measured) is brought to
coincide with the unit vector $\underline{n}_{iq}$ defining the orientation of
the bond $iq$ (see Fig.1e for illustration of the special case where $iq$ and $ij$ lie in the $xz$ plane). This trick reduces the number of
variables in the problem: instead of dealing with two sets of angles,
$\{\varphi_{ij},\theta_{ij}\}$ and $\{\varphi_{iq},\theta_{iq}\}$, we need to
consider only one set $\{\tilde{\varphi}_{ij},\tilde{\theta}_{ij}\}$, which
gives the orientation of the bond $ij$ in the rotated frame. Upon suitably
defining the rotation matrix, the above integral is much simplified.

The rotation is defined around an axis $\underline{t}$ (parallel to
$\underline{e}_{y}$ in the special case of $\phi_{ij}=\phi_{iq}=0$ illustrated in Fig.1e), and perpendicular to both $\underline{e}_{z}$ and
$\underline{n}_{iq}$, with an angle of $\theta_{iq}$ (usual convention of
rotation: counter
clockwise if axis vector points in the direction of the viewer). Here, $\underline{e}_{y}$ and $\underline{e}_{z}$
denote the unit vectors along the $y$ and $z$ axis, respectively, of the Cartesian frame centered on particle $i$. 
Therefore, the
unit vector
\underline{t} defining the rotation axis is:
\begin{equation}
\underline{t}=\frac{\underline{e}_{z}\times \underline{n}_{iq}}{\lvert
\underline{e}_{z}\times \underline{n}_{iq}\rvert}=
\begin{pmatrix}
-\textup{sin}\left ( \phi _{iq} \right )\\
\textup{cos}\left ( \phi _{iq} \right )\\
0
\end{pmatrix}.
\label{tdefinition}
\end{equation}

The rotation matrix $\underline{\underline{R}}$ is defined by the Rodrigues'
formula~\cite{rektorys1969survey}
\begin{equation}
\underline{\underline{R}}=\textup{cos}\left(\theta_{iq} \right
)\underline{\underline{1}}+\textup{sin}\left( \theta_{iq}\right )
\left [\underline{\underline{t}} \right ]_{\times}+\left ( 1- \textup{cos}\left(\theta_{iq} \right
)\right )\underline{t}\otimes \underline{t}
\label{rotationmatrixdefinition}
\end{equation}
where $\underline{\underline{1}}$ represents the identity matrix. Further, we
defined
\begin{equation}
\begin{aligned}[t]
\left [ \underline{\underline{t}} \right ]_{\times}&=
\begin{pmatrix}
0 & -t_{z} & t_{y} \\
t_{z} & 0 & -t_{x} \\
-t_{y} & t_{x} & 0
\end{pmatrix}\\&=
\begin{pmatrix}
0 & 0 & \textup{cos}\left ( \phi_{iq} \right) \\
0 & 0 & \textup{sin}\left ( \phi_{iq} \right)\\
-\textup{cos}\left ( \phi_{iq} \right) & -\textup{sin}\left ( \phi_{iq} \right)
& 0
\end{pmatrix}.
\end{aligned}
\end{equation}

Next, we look at the integral $I_{\alpha\iota\xi}$ defined as:
\begin{equation}
I_{\alpha\iota\xi}=\int_{\Omega-\Omega_{\mathrm{cone}}}n_{ij}^{\alpha
}n_{ij}^{\iota}n_{ij}^{\xi}\textup{sin}\left (\theta_{ij} \right
)\textup{d}\theta_{ij}\textup{d}\phi_{ij}.
\end{equation}
This integral occurs in the expression for $A_{\alpha,\iota \xi \kappa \chi }$,
and considering that $\rho_{ij}(\Omega_{ij}\mid\Omega_{iq})=const$ in the allowed
solid angle $\Omega-\Omega_{\mathrm{cone}}$ for $ij$, we have factored $\rho_{ij}(\Omega_{ij}\mid\Omega_{iq})=const$ out of the $ij$ integral leaving
a product between $I_{\alpha\iota\xi}$ and $\rho_{ij}(\Omega_{ij}\mid\Omega_{iq})$ inside the integral of Eq.5(b),
\begin{equation}
A_{\alpha,\iota \xi \kappa \chi }
=\int_{\Omega}I_{\alpha\iota\xi}\rho_{ij}(\Omega_{ij}\mid\Omega_{iq})\rho_{iq}\left(\Omega_{iq}\right)
n_{iq}^{\alpha}n_{iq}^{\kappa}n_{iq}^{\chi} \textup{d}\Omega_{iq}.
\end{equation}

As is shown in the SI, in the new rotated frame,
one obtains:
\begin{equation}
I_{\alpha\iota\xi}=\int_{\tilde{\theta}_{ij}=\theta_{min}}^{\pi}\int_{\tilde{\phi}_{ij}=0}^{2\pi}n_{ij}^{\alpha
}n_{ij}^{\iota}n_{ij}^{\xi}\textup{sin}\left (\tilde{\theta}_{ij} \right
)\textup{d}\tilde{\theta}_{ij}\textup{d}\tilde{\phi}_{ij}.
\label{transformedintegral}
\end{equation}
$\theta_{min}$
is determined by the excluded volume cone as
$\theta_{min}=2\psi=2\cdot\textup{arcsin}\left( \sigma/2R_{0} \right)$.

We recall that $n_{ij}^{\alpha}$ is defined as the $\alpha$ Cartesian
coordinate of the bond unit
vector $\underline{n}_{ij}$ and is related to the bond unit vector of the
rotated frame
$\underline{n}_{ij,rot}$ via
$\underline{n}_{ij}=\underline{\underline{R}}\cdot\underline{n}_{ij,rot}$, with $\underline{\underline{R}}$ given by Eq.(7).
The bond unit vector in the rotate frame $\underline{n}_{ij,rot}$ is defined by the pair of angles 
${\tilde{\theta}_{ij},\tilde{\phi}_{ij}}$ which represent the integration variables in Eq.(11). 
Therefore, we can now use Eq.(11) together with Eq.(10) to arrive at the
following expression for $A_{\alpha,\iota \xi \kappa \chi }$:
\begin{equation}
\begin{aligned}
A_{\alpha,\iota \xi \kappa \chi }&=
\int_{\theta_{iq}=0}^{\pi}\int_{\phi_{iq}=0}^{2\pi}\int_{\tilde{\theta}_{ij}=2\psi}^{\pi}\int_{\tilde{\phi}_{ij}=0}^{2\pi}
\rho_{ij}\rho_{iq} n_{iq}^{\alpha}n_{iq}^{\kappa}n_{iq}^{\chi}\\
& \times
n_{ij}^{\alpha}n_{ij}^{\iota}n_{ij}^{\xi}\textup{sin}\left(\tilde{\theta}_{ij}\right)\textup{sin}\left(\theta_{iq}\right)
\textup{d}\tilde{\theta}_{ij}\textup{d}\tilde{\phi}_{ij}\textup{d}\theta_{iq}\textup{d}\phi_{iq}.
\end{aligned}
\end{equation}
With the last Eq.(12), we have reduced the original integral for
$A_{\alpha,\iota \xi \kappa \chi }$ to a much simpler integral with
well-defined integration limits in the solid angle. The integral can be easily
evaluated using $\rho_{iq}=1/4\pi$, which accounts for the fact that the
orientation of $iq$ can be freely chosen, whereas
$\rho_{ij}(\Omega_{ij}\mid\Omega_{iq})=1/(4\pi-\Omega_\mathrm{cone})=1/3\pi$ due to the restriction imposed
by excluded-volume.

From the evaluation of the integral we obtain the following numerical values of the coefficients,
\begin{equation}
\begin{tabular}{c||c|c|c|}
$\alpha$                & $x$           &        $y$    &       $z$ \\ \hline
$A_{\alpha,xxxx}$       &$-0.0304$      & $-0.00357$&$-0.00357$ \\
$A_{\alpha,xyxy}$       &$-0.00357$     & $-0.00357$&$-0.000149$\\
$A_{\alpha,xxyy}$       & $-0.00982$& $-0.00982$&$-0.00327$\\
 \end{tabular}
\end{equation}
We also recall that $B_{x,xxxx}=1/7$, $B_{y,xxxx}=1/35$, $B_{z,xxxx}=1/35$
$B_{x,xyxy}=B_{y,xyxy}=1/35$, $B_{x,xxyy}=B_{y,xxyy}=1/35$, $B_{z,xyxy}=B_{z,xxyy}1/105$ as obtained in Ref.~\cite{zaccone2011approximate}.
Using these values of coefficients in Eq.(3), for
shear in the $xy$ plane we find: $G=(1/30)\kappa
R_{0}^{2}(N/V)(z-z_\mathrm{iso})+G_\mathrm{corr}$, where $z_\mathrm{iso}=2d=6$
and the correction term due to excluded-volume correlations is $G_\mathrm{corr} = 0.0218$, in units of $\kappa
R_{0}^{2}(N/V)$.
The anisotropy of the shear field leaves a small projection of the
interparticle forces in the direction of the opposing bonds, which leaves
nonaffinity nearly intact under shear.

\textbf{Discussion}\\
The non-zero, though small, $G_{corr}$ predicted by the analytical theory might
be due to model approximations
which are intrinsically different from approximations and assumptions done in
numerical simulations. For example, we
always overestimate the excluded-volume cone by not considering the
deformability of the soft particles in jammed packings. If this was
properly taken into account, it would lead to a smaller excluded-volume cone
and weaker correlations, hence to a higher nonaffinity than predicted in this
approximation. 
In turn, that would yield
an even smaller, practically negligible, value of
$G_\mathrm{corr}$. Another, though related, source of inaccuracy is the neglect
of deviations from the average nearest-neighbour
distance $R_{0}$. These deviations are possible if the particles are allowed to
deform slightly at contact. There are also other differences in terms of boundary conditions and the structure of the packing cannot obviously be exactly the same for theory and simulations. Further, we do not take into account 
\textit{local} chemistry-related effects at the interface between grains/drops
(which may control how the creation of excess contacts $z-z_{c}$ depends upon
$\phi$ under different physico-chemical conditions~\cite{weitz1,weitz2,wyart}).
This is so because we want to focus on the more general many-body physics which
controls the mechanical deformation behaviour (i.e. how $G$ and $K$ vary with
$z$).

In a similar way, for the bulk modulus we obtain
$K=(1/18)\kappa\sigma^{2}(N/V)(z-z_\mathrm{iso})+K_\mathrm{corr}$. In this case
$K_\mathrm{corr}=0.087$, always in units of $\kappa
R_{0}^{2}(N/V)$, is significantly larger. The reason why $K_\mathrm{corr} \approx 4 G_\mathrm{corr}$
lies in
the fact that the forces transmitted by neighbours are on average cancelling each other effectively
under isotropic compression, though not to the same extent in shear. The latter is
strongly anisotropic and causes the forces transmitted by 
neighbours to be misaligned such that the cancellation of nearest-neighbour forces with same orientation and opposite direction is not as effective. Our theoretical predictions match the known effect of vanishing
of the ratio $G/K$ at the rigidity transition \cite{ellenbroek2009non}
$z_\mathrm{iso}=2d=6$.
The analysis for the centrosymmetric crystal based on the affine assumption can
be found in
Born's work and gives the constant ratio \cite{Born1954} $G/K=0.6$, independent
of
$z$. The same ratio is also found in the simulations of Ref.~\cite{goodrich}. This
limit is captured by our general framework of disordered lattice dynamics, as
both sets of coefficients $A_{\alpha,\iota \xi \kappa \chi }$ and
$B_{\alpha,\iota \xi \kappa \chi }$ are identically zero for centrosymmetric
crystals, giving $G_\mathrm{NA}=0$ and $K_\mathrm{NA}=0$.

We have seen above that the shear modulus does not completely vanish at the
isostatic transition, but remains small and equal to $G_{corr}=0.0218$, and that
the ratio  $G_{corr}/K_{corr}$ is about 0.26. Hence, our theory gives an order
of magnitude $\mathit{O}(10^{-1})$, instead of $\mathit{O}(0)$, as many
numerical simulations seem to suggest upon extrapolation to $z=z_{c}$.
On the other hand, however, our theory is the only analytical approach which
predicts a substantial difference, close to one order of magnitude, between $K$
and $G$. In many amorphous and other non-centrosymmetric materials, the
difference between shear and bulk modulus is about a factor $4$, like for
example in crystalline ice and quartz~\cite{Helmreich,Bechmann}, which is very
consistent with our result.

\begin{figure*}[t]
\centering
{\includegraphics[width=0.95\textwidth]{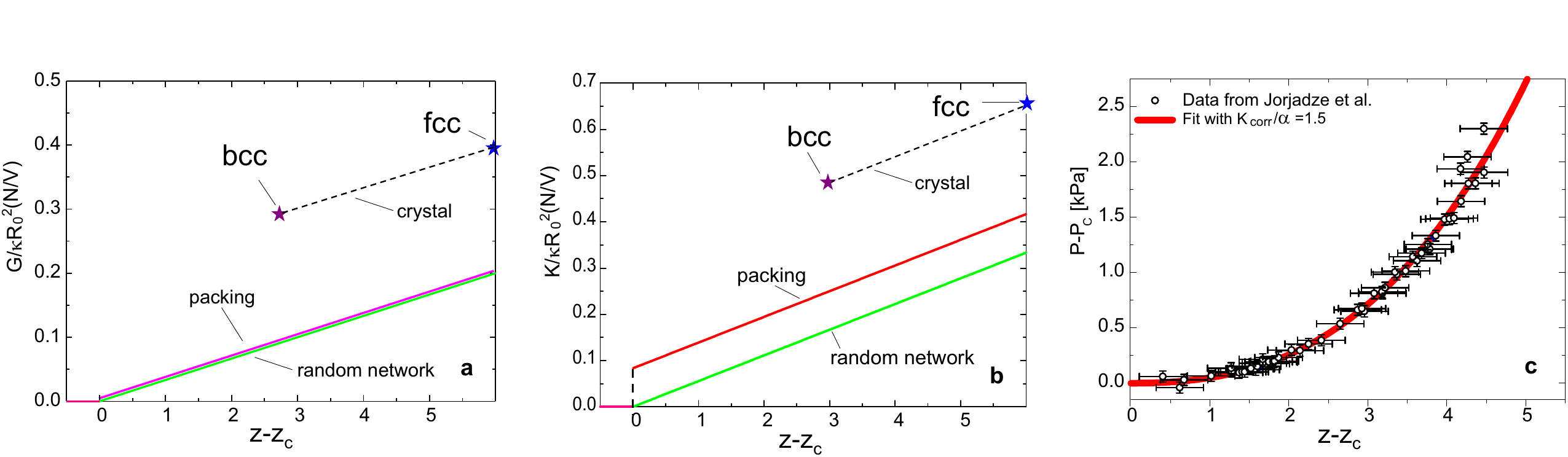}
\caption{\textbf{Theoretical predictions in different limits across the
disorder spectrum:} \
(a) Theoretical predictions for the shear modulus $G$ near the isostatic limit
$z \geq z_\mathrm{iso}$, for crystals, jammed packings and random networks. The small term $G_{corr}=0.0218$ which contributes to the packing shear modulus has been neglected in line with the considerations presented in the text. (b)
Similar predictions for the bulk modulus $K$ for crystals, jammed packings and
random networks, where now $K_{corr}$ is making an important contribution to the packing bulk modulus. (c) Fit of experimental data  of Ref.\cite{jorjadze2013microscopic} on compressed emulsion, using our Eq.(14)
with the only fitting parameter given by $\alpha \approx 0.17~kPa$. }}
\label{fig2}
\end{figure*}

Our theoretical predictions are presented in Fig.2a,b for the shear and the
bulk moduli, respectively. It is evident that the random network is the overall
softest system because even if the shear modulus is basically the same as for the jammed
packing (apart from the relatively small term $G_{corr}=0.0218$ in the packing modulus which we neglected in the plot), its bulk modulus is significantly smaller. The reason is that the bulk
modulus of the packing behaves closer to the affine deformation limit due to
the reduction of nonaffinity caused by excluded-volume correlations, as
explained above. Intriguingly, the same behaviour (soft shear modulus,
quasi-affine bulk modulus)
is well known to occur in atomic amorphous materials, such as amorphous
Gallium~\cite{Kinder}.
In the random network, instead, the nonaffinity is strongest
because no cancellation of forces due to local particle correlations can occur.
This microscopic mechanism thus explains what observed in recent numerical
simulations where
this difference between packings and networks was investigated
numerically~\cite{ellenbroek2009non}.
What was interpreted as an "anomalous" behaviour, can be explained
mechanistically based on nonaffinity.

Finally, our microscopic theory provides a quantitative prediction of moduli
and of the discontinuous jump of the bulk modulus at the jamming transition,
quantified by $K_\mathrm{corr}$. We introduce the shorthand
$\beta=(1/30)\kappa\sigma^{2}(N/V)$ and
$\alpha=(1/18)\kappa\sigma^{2}(N/V)$ for the prefactors of $G$ and $K$, respectively, for convenience of notation.
Recalling that $\kappa$ has units of $N/m$, $\sigma$ is a length and $N/V$ is in units of $m^{-3}$, it is clear that $\alpha$ and $\beta$ are measured in units of $Pa$, although here we discuss their calculated values in units of $\kappa\sigma^{2}(N/V)$.
Calculating the slope  $G \approx
\beta(z-z_\mathrm{iso})$, we find $\beta \approx 0.60$ for the shear modulus,
in good agreement with the value $\beta \approx 0.75$ found in the simulations
of Goodrich et al.\cite{goodrich}.
For the jump in the bulk modulus at jamming, using the short-hand $K \approx
\alpha(z-z_\mathrm{iso})+K_{corr}$, our theory gives $K_{corr}/\alpha=1.50$,
which is of the right order of magnitude but smaller than the value
$K_{corr}/\alpha=4.50$ given by Goodrich et al.\cite{goodrich}. This
discrepancy
might be due to the obviously different approximations and assumptions done in
numerical simulation protocols, which were discussed at the beginning of this section.

\textbf{Comparison with compressed emulsions}\\
We also compared our prediction for the jump of compressibility with recent
experiments on compressed emulsions~\cite{jorjadze2013microscopic}. In the
experiment, different values of pressure applied to the packing were recorded,
and the values of $z$ corresponding to the different pressure values were
measured using a fluorescent dye in the interparticle contacts between emulsion
droplets. The output of this measurement is a curve relating $\delta P=P-P_{c}$
to $\delta z=z-z_{c}$, where we have to interpret $z_c$ as the limit of
isostaticity. The bulk modulus is defined in terms of pressure and coordination
$z$ via $K=-V (\textup{d}P/\textup{d}V)=-V(\textup{d}P/\textup{d} \delta
z){\textup{d} \delta z}/{\textup{d}V}$. There is a one-to-one mapping between
the volume fraction occupied by the drops, $\phi$, and the contact number, $z$,
in compressed emulsions, which was determined
empirically in Ref.\cite{jorjadze2013microscopic} to be
$\delta z=z_{0} \sqrt{\delta \phi}$, with $z_{0}=10.6$, for their system. Using
this
relation, and the definition of volume fraction $\phi=V_\mathrm{drops}/V$, one obtains:
${\textup{d} \delta z}/{\textup{d}V}=-{z_{0}V_\mathrm{drops}}/{2\sqrt{\delta
\phi}V^{2}}=- {z_{0}^{2}\phi}/{2 \delta z V}$. Upon replacing in the formula
for $K$, we finally have a relationship between $K$, $\delta z$, and $\delta
P$, given by
$K={\phi z_{0}^{2}}/{2 \delta z} ({\textup{d}P}/{\textup{d} \delta z})$. We can
thus replace our theoretical expression for $K=\alpha \phi \delta z
+K_\mathrm{corr}$ where $\alpha$ is the only fitting parameter containing the
spring constant, and integrate the differential equation to get
\begin{equation}
\delta P=P-P_{c}=\frac{K_{corr}}{z_{0}^{2}}(\delta z)^{2}
+\frac{2\alpha}{3z_{0}^{2}}(\delta z)^{3}. \label{pressure}
\end{equation}

The one-parameter fit comparison between the analytical theory, given by Eq.(14)
and the experimental data of Ref.\cite{jorjadze2013microscopic} is shown in
Fig.2c. The only fitting parameter is $\alpha \propto \kappa/R_{0}$ which is
directly proportional to the spring constant of the drop-drop interaction,
hence contains the dependence on the particular chemistry of the emulsion, and
inversely proportional to the drop diameter.
Our fitting accounts for both creation of excess contacts with pressure, and
nonaffine
particle rearrangements, and is able to provide a one-parameter fit of the
data.
In Ref.\cite{jorjadze2013microscopic} the same data were modelled by accounting
for the creation of
excess contacts only, and neglecting rearrangements, which requires two
adjustable parameters.
Hence, a more quantitative description of experimental data can be achieved
using the new framework proposed here.

\textbf{Conclusions}\\
We showed that
the mechanical response of solids is strongly affected by the degree of local
orientational order of the
lattice, whether fully enforced (as in centro-symmetric crystals), low (as in
random
networks), or intermediate due to excluded-volume constraints in jammed
packings). In particular, intermediate degrees of orientational order are very
relevant for
amorphous solids as documented by numerical simulations and experiments (see
e.g. Refs.~\cite{Tanaka2010,Tanaka2013}).
Our theory shows that
the lower the local orientational order,
the stronger is the role of internal nonaffine deformations which always soften
the mechanical response.
With excluded-volume correlations, as in packings, there is significant local
orientational order~\cite{Tanaka2013} and two bonds
can have the same orientation across a common neighbour, due to
excluded-volume correlations. The forces transmitted by these
nearest-neighbours cancel each other completely under compression, thus considerably
reducing nonaffinity and softening for the compression mode.
For lattices with strong excluded-volume like random packings (but also atomic
materials like amorphous Gallium),
our theory predicts that the bulk modulus can be a factor of 4 larger than the
shear modulus, which is in semi-quantitative or at least qualitative agreement
with both simulations~\cite{ellenbroek2009non,goodrich} and experiments on
atomic~\cite{Kinder} and molecular materials~\cite{Helmreich}.
Furthermore, our theory provides an excellent quantitative description of the
dependence
of the bulk modulus of compressed emulsions on the microscopic coordination
number, with just one fitting parameter in the comparison with experiments~\cite{jorjadze2013microscopic}.
We also expect that our lattice dynamics framework for materials
that lack inversion symmetry can lead to a better understanding of the role of
phonon lattice instabilities on the critical temperature of
superconductors~\cite{sigrist}.\\

\begin{acknowledgments}
This work was supported by the Theoretical Condensed Matter programme grant
from EPSRC.
M.S. thanks the Konrad-Adenauer-Stiftung for their financial support.
\end{acknowledgments}

\textbf{Author contribution statement}\\
M.S. and A.Z. developed the theory and the calculations, A.Z. and E.M.T. designed the research and J.B.
provided the experimental context. A.Z. wrote the manuscript with the collaboration of E.M.T. A.Z., E.M.T., and J.B. reviewed the manuscript.

\textbf{Competing financial interests}\\
The authors declare no competing financial interests.

\end{document}